# Circuit Analysis in Metal-Optics

**M. Staffaroni[1], J. Conway[2], S. Vedantam[2], J. Tang[2], E. Yablonovitch[1]**

[1] UC Berkeley EECS Department, Berkeley, CA, 94720

[2] UCLA Electrical Engineering Dept., Los Angeles, CA, 90095

ABSTRACT

We provide electrical circuit descriptions for bulk plasmons, single surface plasmons, and parallel-plate plasmons. Simple circuits can reproduce the exactly known frequency versus wave-vector dispersion relations for all these cases, with reasonable accuracy. The circuit paradigm directly provides a characteristic wave-impedance, $Z_o$, that is rarely discussed in the context of plasmonics. The case of a single-surface-plasmon is particularly interesting since it can be modeled as a transmission line, even though there is *no* return current conductor. The capacitance/unit length and the Faraday inductance/unit length, of a flat metal surface, are $C'=2\varepsilon_o kW$, and $L'=\mu_o/2kW$ respectively, (where k is the wave-vector, and W is the width of the flat metal surface). We believe that many other metal-optic geometries can be described within the circuit paradigm, with the prerequisite that the distributed capacitance and inductance must be calculated for each particular geometry.

**I. Introduction**

Circuits with distributed inductive and capacitive elements can capture much of the physics in Maxwell's Equations. A circuit model provides powerful insights, and can reveal physics that might otherwise be concealed within an exact analytical solution, or in a brute-force numerical solution.

While lumped element circuit approaches are common in the microwave and RF regime, they have played a limited role in optics. Lumped optical circuit models are now available for metallic nano-spheres[1,2] and split-ring resonators[3,4], but the simplest case of the electrical circuit for a flat metallic surface plasmon has not been presented.

In addition to the capacitance, and Faraday inductance, there is also kinetic inductance arising from the inertia of the electrons in a metal. Kinetic inductance dominates over Faraday inductance at blue frequencies, or when there are characteristic dimensions smaller than the collisionless skin depth[5]. When kinetic inductance dominates over Faraday inductance, that is properly called the plasmonic regime[6]. In general, all three circuit components, must be included, capacitance, Faraday inductance, and kinetic inductance.

In this article we model guided wave propagation in metal optics as a 1-dimensional distributed element circuit. Simple circuit models do recover key results pertaining to dispersion curves for: *(i)* guided surface plasmon waves on a single metal surface, and *(ii)* for parallel plate optical waveguides. The circuit approach also provides new insights into metal optics that would be lost in more rigorous formal or numerical treatments. Namely wave impedance emerges, and is recognized as of equal significance to plasmon dispersion. Owing to the presence of kinetic inductance, a plasmonic transmission line can surprisingly have an impedance greater than the impedance of free space, $Z_o = \sqrt{(\mu_o/\varepsilon_o)} = 377\Omega$, (where $\mu_o$ & $\varepsilon_o$ are the permeability and permittivity of free space, respectively). The ability to adjust the plasmon wave impedance allows voltage transformer action at optical frequencies, through tapered metallic structures.

This transformer action is a unique feature of plasmonics and can be used:

*(i)* to engineer efficient delivery of optical power to the nanoscale, or

*(ii )* as an impedance matching tool toward molecular light emitters.

## II. Bulk Plasma Resonance condition

The bulk plasmon resonance condition may be derived from Newton's $2^{nd}$ Law, F=ma. Consider a rectangular metal slab of length z, and cross-sectional area, A, as shown in Fig. 1.

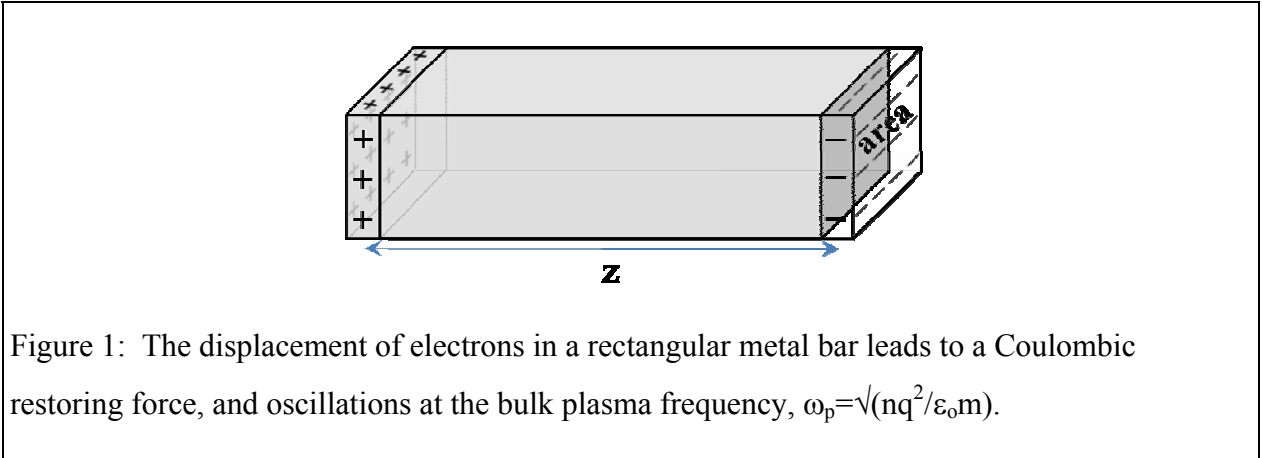

Figure 1: The displacement of electrons in a rectangular metal bar leads to a Coulombic restoring force, and oscillations at the bulk plasma frequency, $\omega_p = \sqrt{(nq^2/\varepsilon_o m)}$.

Displacing the electron cloud about the background ionic lattice by an infinitesimal distance dz along the length dimension creates an electric field V/z. The force on each electron is then F=qV/z = ma=m(dv/dt). It is wise then, to insert the quantity nqA, in both numerator and denominator

$$F = qV/z = ma = m\frac{dv}{dt} = m\frac{nqA}{nqA}\frac{dv}{dt}, \quad \ldots\ldots\ldots\ldots\ldots\ldots\ldots\ldots\ldots\ldots\ldots\ldots\ldots\ldots(1)$$

where n is the number density of electrons in the metal, and q is the electric charge. Recognizing that nqAv is the electric charge that passes a single point in one second, it represents the flowing electric current, nqAv=I. Eq'n. (1) can then be re-written as qV/z = (m/nqA)(dI/dt) which resembles the voltage response of an inductor: $V = (m/nq^2)(z/area) \times (dI/dt)$, where the inductance is $L_{kinetic}=(m/nq^2)(z/area)$. Likewise the charge separation in the metal slab leads to a capacitance: $C=\varepsilon_o(area/z)$. When the charge cloud is released, it oscillates about the ionic lattice like a mass on a spring, like a plasmon[7], or simply as an LC circuit. The corresponding plasma oscillation frequency is given by:

$$\omega^2 = \frac{1}{LC} = \frac{nq^2}{\varepsilon_o m}\frac{z}{area}\frac{area}{z} = \frac{nq^2}{\varepsilon_o m} \equiv \omega_p^2 \quad \ldots\ldots\ldots\ldots\ldots\ldots\ldots\ldots\ldots(2)$$

Thus bulk plasmons can be represented by a kinetic inductance $L_{kinetic}=(m/nq^2)(z/area)$. Kinetic inductance is a well-known concept in superconductitivity[8], where the ordinary resistance is zero, and it is only inductance that impedes current flow.

### III. Equivalence Between Metal Dielectric Constant & Conductivity

At the root of the circuit description of metal optics, is a recognition of the equivalent parameterization between optical dielectric constant of a metal, $\varepsilon_m(\omega)$, and the less frequently discussed complex frequency dependent optical conductivity $\sigma(\omega) \equiv j\omega\varepsilon_o(\varepsilon_m-1)$. This can be seen directly from Ampere's Law, which can treat the metal as a dielectric:

$$\nabla \times H - \partial D/\partial t = \nabla \times H - j\omega\varepsilon_m\varepsilon_o E = 0, \quad \ldots\ldots\ldots\ldots\ldots\ldots\ldots\ldots\ldots(3)$$

with relative dielectric constant $\varepsilon_m$. Alternately the metallic response can be regarded as producing only currents J and charges $\rho$ with otherwise negligible dielectric response, in which case Ampere's Law becomes:

$$\nabla \times H - j\omega\varepsilon_o E = J = \sigma E = j\omega(\varepsilon_m-1)\varepsilon_o E \quad \ldots\ldots\ldots\ldots\ldots\ldots\ldots\ldots\ldots(4)$$

The equivalence of eq'ns: (3)&(4) is ensured when the complex conductivity is defined as $\sigma(\omega) \equiv j\omega\varepsilon_o(\varepsilon_m-1)$. The complex resistivity $\rho(\omega) \equiv 1/\sigma(\omega)$ can be rationalized into real and imaginary parts as $\rho(\omega) \equiv \frac{1}{\sigma(\omega)} = \frac{1}{\varepsilon_o \omega} \frac{j(1-\varepsilon'_m) + \varepsilon''_m}{(1-\varepsilon'_m)^2 + (\varepsilon''_m)^2}$. Then a metallic wire will have an

impedance that can be derived from the complex resistivity:

$$Z = \frac{1}{\sigma(\omega)} \frac{\text{length}}{\text{area}} = \frac{1}{j\omega\varepsilon_o(\varepsilon_m - 1)} \times \frac{\text{length}}{\text{area}} \quad \text{..............................(5)}$$

Substituting in the metal optical relative dielectric constant; $\varepsilon_m = 1 - nq^2/\varepsilon_o m\omega^2$, that neglects collisions, the impedance becomes $Z = j\omega \times (m/nq^2) \times (\text{length/area}) \equiv j\omega L_{kinetic}$. Thus the kinetic inductance arises from the inertia, m, of the $L_{kinetic} = (m/nq^2) \times (\text{length/area})$, as in the previous section on bulk plasmons. Equivalently expressed in terms of relative dielectric constant $L_{kinetic} = (1/\omega^2\varepsilon_o(1-\varepsilon_m)) \times (\text{length/area})$. Including collisions, there is an additional resistance term $Z = R + j\omega L_{kinetic}$, where for collision time $\tau$, $R = (m/nq^2\tau) \times (\text{length/area})$, which is the usual expression for the dissipative resistance of an electron gas.

*IV. Circuit Theory for Surface Plasmons*

A hallmark of metal-optics is that the interface between a metal and free space can support surface plasmon modes[9,10,11] with the exact dispersion relation:

$$k = \frac{\omega}{c}\sqrt{\frac{\varepsilon_m}{\varepsilon_m + 1}} \quad \text{...................................(6)}$$

where $k = 2\pi/\lambda_k$ is the wave vector of the mode, and $\lambda_k$ is the corresponding wavelength along the surface. Eq'n. (6) is plotted as the solid blue curve in Figure 2. But the physics of surface plasmons can be captured in a distributed circuit transmission line model that is quite conventional, except that it includes $L_{kinetic}$ in series with the more conventional Faraday inductance, $L_F$. For example in the circuit model, at large wave-vectors, $\omega$ has the constant value $\omega_p/\sqrt{2}$, or equivalently $\varepsilon_m(\omega) = -1$, while at small wave-vectors $\omega = kc$.

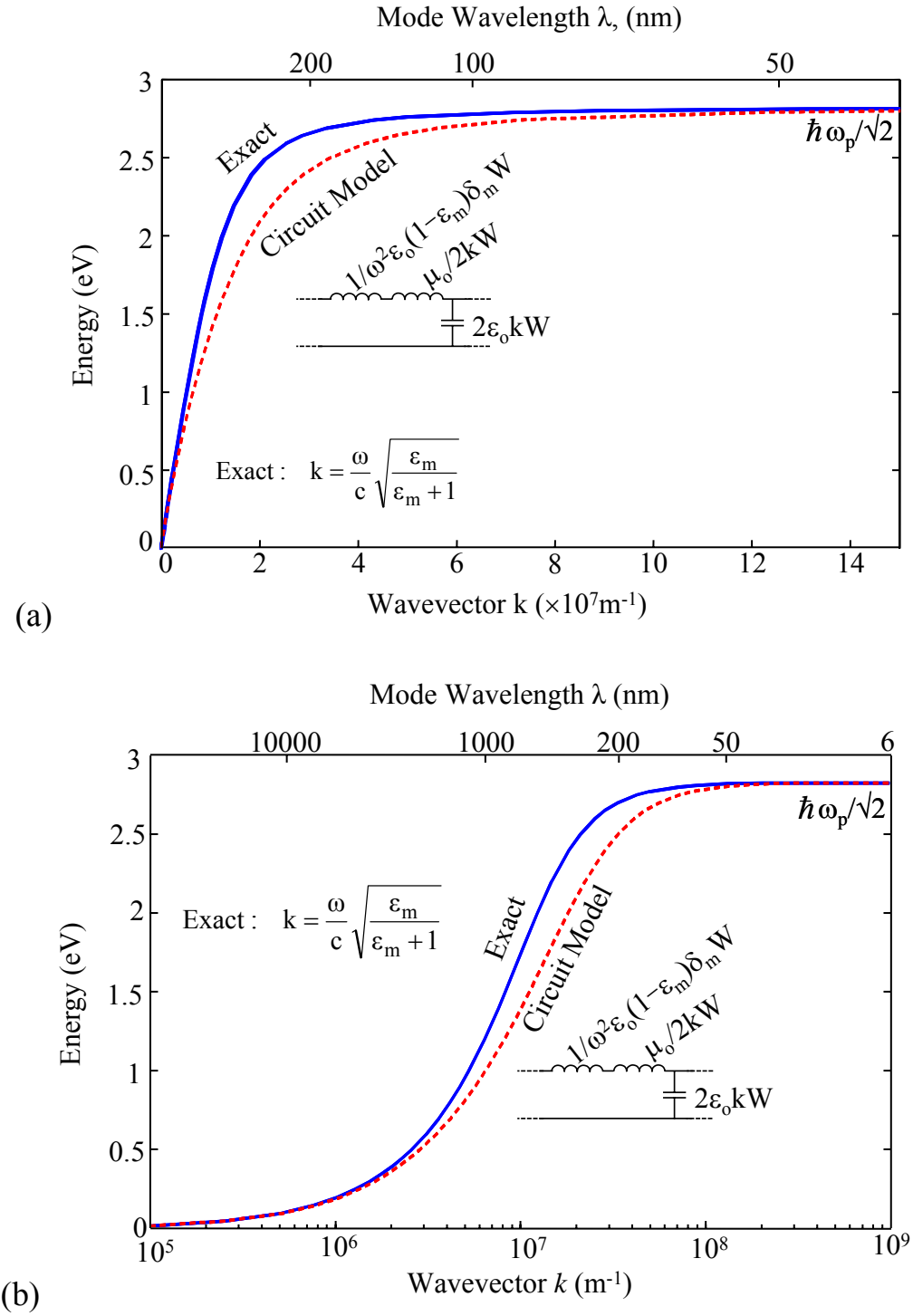

Figure 2: The dispersion relation of surface plasmons for a Drude metal with $\hbar\omega_p = 4$ eV. (a) & (b) are linear and semi-logarithmic plots respectively. The solid blue curve is the exact dispersion, while the red, dashed line is the transmission line circuit model consisting of capacitance C', kinetic inductance $L'_k$, and ordinary Faraday inductance, $L'_F$., where the prime ' represents--per unit length of transmission line.

In developing a transmission line circuit model, there are several challenges to overcome: Is it indeed possible to have a transmission line when there is only a single conductor? Transmission line theory usually applies when there are two conductors, one to transmit current, and one to return current. We want to treat the single metal surface of Fig. (3) as a conductor, but there is no return conductor to complete the circuit! Effectively the return currents must flow at infinity, to permit a single metal plate to act as a transmission line.

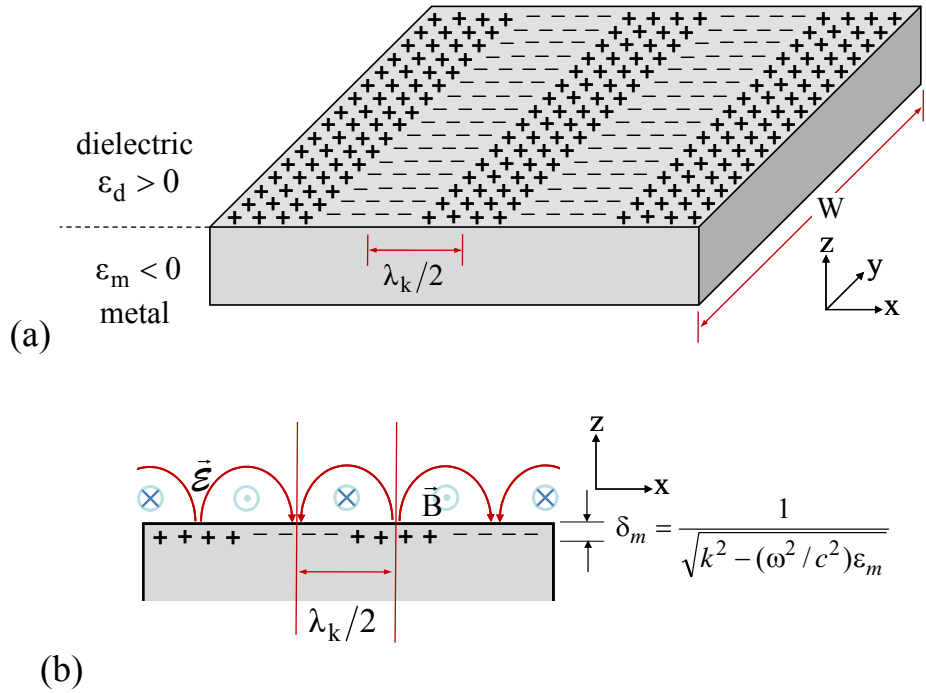

Figure 3: (a) The distribution of charges associated with a surface electromagnetic wave on a metal. (b) The electric and magnetic fields, and the associated skin depth in which metallic current flows.

The charge distribution $\sigma(x)$ in Fig. 3(a) creates an oscillatory voltage $V(x)$ which can be calculated from electrostatics. With voltage and charge calculated, a capacitance per unit length $C'$ in the x-propagation direction can be determined. A detailed electrostatic calculation, in appendix A, shows that $C'=2\varepsilon_o kW$, where W is the width of the metal surface in the y-direction, transverse to $k_x$, the wave propagation direction.

As the surface charge shown in Fig. 3(a) oscillates in time, sinusoidal surface currents must flow in space. The surface currents produce a magnetic field B(x,z) above the surface, spatially sinusoidal in the x-direction. The effective Faraday inductance $L_F$ can be calculated[12] from $\int B\, dxdz = L_F I$, where I is the surface current over the full metal width W, and the magnetic flux is obtained by integrating $\int B dxdz$ above the metal surface in the +z-direction, and in the x-direction. Expressed as inductance/per unit length $L'_F$ in the x-direction, the magnetic flux is $\int B\, dxdz = \int L'_F I\, dx$. Equating the integrands yields $L'_F I = (\int B\, dz)$. Thus a calculation of $(\int B dz)/I$, allows us to determine $L'_F$, the Faraday inductance per unit length. In appendix B, it is shown that $L'_F = \mu_o/2kW$.

There remains to calculate the contribution from the kinetic inductance, $L_k = (1/\omega^2 \varepsilon_o(1-\varepsilon_m))\times(\text{length/area})$ which is generally in series with $L_F$. Per unit length $L'_k = (1/\omega^2 \varepsilon_o(1-\varepsilon_m))\times(1/\text{area})$. It remains to calculate the area of the conduction path, which is the skin depth of the metal slab × the width, $A=\delta_m W$.

There is no one skin depth that is appropriate to all situations. Table I presents three different forms of skin depth:

| Skin Depth Type | (a) Collisional $\omega\tau<1$ | (b) Collisionless $\omega\tau>1$ | (c) Surface Wave Skin Depth |
|---|---|---|---|
| Skin Depth; $\delta_m$ | $\sqrt{\dfrac{2\rho'(\omega)}{\omega\mu_o}} = \dfrac{c\sqrt{2\varepsilon''_m(\omega)}}{\omega\|1-\varepsilon_m\|}$ | $\dfrac{c}{\omega\sqrt{1-\varepsilon'_m}} = \dfrac{c}{\omega_p}$ | $\dfrac{1}{\sqrt{k^2 - \varepsilon_m(\omega^2/c^2)}}$ |

Table I: (a) Collisional skin depth is appropriate to microwaves, and is found in most Electromagnetics books[13]. (b) Collisionless skin depth[5] pertains to normal incidence plane waves, above the collision frequency, ~10THz.. (c) If there is a wave propagation k, parallel to the surface, the surface wave skin depth applies[9]. Generally $\varepsilon_m$ is predominantly negative, and $\varepsilon'_m$ & $\varepsilon''_m$, and $\rho'$ & $\rho''$, represent the real and imaginary parts respectively.

For the surface wave propagation that we are considering, the appropriate skin depth $\delta_m$ is given as case (c) of Table I. Thus $L'_k = 1/\omega^2\varepsilon_o(1-\varepsilon_m)\delta_m W$.

All three component values, $C'$, $L'_F$, and $L'_k$ are now known. Since they are defined per unit length, they contribute toward a distributed transmission line. The two inductors $L'_F$, and $L'_k$ act in series $L'_F+L'_k$, as shown in the inset of Figure 2. The properties of such a series L,--parallel C transmission lines are well worked out[14]. There are two important properties of transmission lines;

(a) The dispersion, $\omega$ versus k, given by $\omega^2=k^2/L'C'$,

(b) The wave impedance given by $Z=\sqrt{(L'/C')}$.

The general dispersion is given by $\omega^2=k^2/(L'_F+L'_k)C'$ which can be rewritten:

$$\frac{1}{\omega^2} = \frac{1}{k^2}\left(\frac{\mu_o}{2kW} + \frac{1}{\omega^2\varepsilon_o(1-\varepsilon_m)\delta_m W}\right)2\varepsilon_o kW = \frac{1}{k^2}\left(\mu_o\varepsilon_o + \frac{2k}{\omega^2(1-\varepsilon_m)\delta_m}\right) \quad\ldots\ldots\ldots(7)$$

Equation (7) is plotted as the "circuit model" in Fig. 2.

In the limit $k<\omega_p/c$, the kinetic inductance term is negligible compared to the Faraday inductance. Eq'n: (7) simplifies to $\omega=kc$, the "light line", in good agreement with the exact solution in Fig. 2.

In the limit $k>\omega_p/c$, the kinetic inductance term dominates the Faraday inductance. Moreover, the skin depth $\delta_m=1/\sqrt{(k^2-\varepsilon_m\omega^2/c^2)}$, becomes $\delta_m\approx 1/k$. Then Eq'n. (7) becomes:

$$\frac{1}{\omega^2} = \frac{1}{k^2}\left(\frac{2k^2}{\omega^2(1-\varepsilon_m)\delta_m}\right) = \frac{2}{\omega^2(1-\varepsilon_m)} \quad\ldots\ldots\ldots\ldots\ldots\ldots\ldots\ldots\ldots(8)$$

Which further reduces to $1-\varepsilon_m=2$, or in other words $\varepsilon_m=-1$, which can also be written $\omega_p=1/\sqrt{2}$, all of which expressions correspond exactly to the surface plasmon condition, in good agreement with the exact dispersion in Fig. 2.

In the intermediate regime $k \sim \omega_p/c$, the circuit model deviates from the exact dispersion in Fig. 2 by about 15%. The circuit model is distributed, consisting repeating circuit blocks in one-dimension, as shown in the insets of Fig. 2. A more realistic model would be distributed in two-dimensions respecting the 2 dimensional character of our problem. A 2-dimensional distributed circuit would better describe our situation, but such an avenue would add many more circuit components while doing little for intuitive understanding. Thus we retain our 1-dimensionally distributed model in spite of the slight disagreement with the exact solution.

The wave impedance is of equal importance to the dispersion, and in the circuit model may be written:

$$Z = \sqrt{\frac{L'}{C'}} = \sqrt{\frac{\frac{\mu_o}{2kW} + \frac{1}{\omega^2 \varepsilon_o (1-\varepsilon_m) \delta_m W}}{2\varepsilon_o kW}} = \frac{1}{W}\sqrt{\frac{\frac{\mu_o}{2k} + \frac{1}{\omega^2 \varepsilon_o (1-\varepsilon_m) \delta_m}}{2\varepsilon_o k}} \quad \ldots\ldots\ldots(9)$$

The final expression in Eq'n. (9) shows explicitly that the wave impedance $Z \propto (1/W)$ becomes very large as the conducting plate becomes narrower. Once again we treat the limits $k<\omega_p/c$ and $k>\omega_p/c$:

For small wave vectors near the light line, $k<\omega_p/c$, the Faraday inductance dominates:

$$Z = \sqrt{\frac{L'}{C'}} = \frac{1}{W}\sqrt{\frac{\mu_o}{4k^2 \varepsilon_o}} = \frac{1}{2kW}\sqrt{\frac{\mu_o}{\varepsilon_o}} = \frac{1}{2kW} \times 377\Omega \quad \ldots\ldots\ldots\ldots(10)$$

Since $W>1/k$ to maintain the 1-dimensionality of the problem, the impedance in case $k<\omega_p/c$ cannot exceed the impedance of free space, $377\Omega = \sqrt{(\mu_o/\varepsilon_o)}$. In general, for transmission lines without kinetic inductance, $377\Omega$ is an upper limit to the achievable impedance.

In the opposite limit, $k>\omega_p/c$, the kinetic inductance dominates in Eq'n. (9), and wave impedance becomes

$$Z = \sqrt{\frac{L'}{C'}} == \frac{1}{W}\sqrt{\frac{1}{2\omega^2\varepsilon_o^2(1-\varepsilon_m)k\delta_m}} = \frac{1}{W}\sqrt{\frac{\mu_o}{\varepsilon_o}\frac{c^2}{2\omega_p^2 k\delta_m}} = \frac{1}{W}\frac{\lambda_p}{2\pi}\sqrt{\frac{\mu_o}{2\varepsilon_o}} \quad \ldots\ldots\ldots(11)$$

where $c/\omega_p$ was replaced by $\lambda_p$, the vacuum wavelength at the plasma frequency, and $\delta_m \sim 1/k$ in the deep plasmonic regime where kinetic inductance dominates, resulting in the simple form on the right side of Eq'n. (11). When width W is less than the skin depth $\lambda_p/2\pi$, this creates the possibility of a wave impedance $Z>377\Omega$, which should be regarded as a unique feature of the plasmonic regime.

*V. Circuit Theory for the Plasmonic Parallel Plate Waveguide:*

We now transfer our attention to parallel plate waveguides at optical frequencies. There exists an exact solution[10] of Maxwell's equations for the parallel-plate waveguide, for general complex dielectric constant $\varepsilon_m$:

$$\exp(-K_i d) = \frac{K_i \varepsilon_m + K_m}{K_i \varepsilon_m - K_m} \quad \ldots\ldots\ldots\ldots\ldots\ldots\ldots\ldots\ldots\ldots(12)$$

With $K_m = \sqrt{k^2 - \varepsilon_m(\omega/c)^2} = 1/\delta_m$, $K_i = \sqrt{k^2 - (\omega/c)^2}$, and where $k=2\pi/\lambda_k$ is the actual wave vector of the mode, and $\lambda_k$ is the corresponding mode wavelength, $\omega$ is the optical frequency, c is the speed of light in free space, and d is the plate spacing. The skin depth $K_m=1/\delta_m$ in the metal is the same as for the single plate waveguide described in Table I.

In transmission line theory, parallel plate waveguides operating in the microwave regime are modeled as a distributed-element repeating circuit of series inductors and parallel capacitors[12]. The voltage and current waveforms supported by this type of reactive transmission line circuit follow the general dispersion relation given by $\omega^2=k^2/L'C'$, and wave impedance by $Z=\sqrt{(L'/C')}$. Once again, we need to define $L'_F$, $C'$, and $L'_{kinetic}$. For the parallel plate geometry the kinetic inductance $L'_k$ is the same as for a single plate, but multiplied by 2 to account for the series inductance of the first plate and the return current plate, $L'_k = 2/\omega^2\varepsilon_o(1-\varepsilon_m)\delta_m W$.

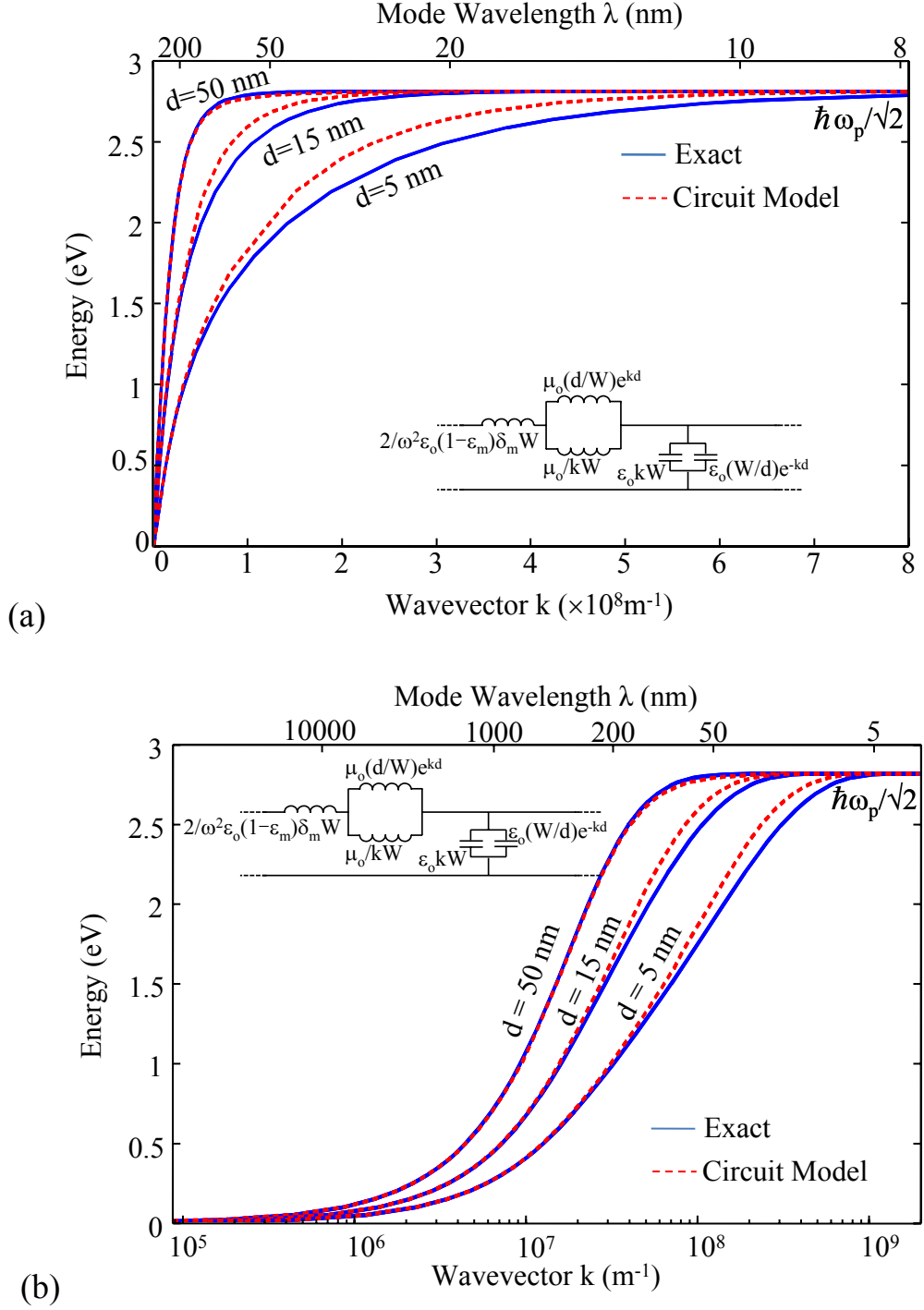

Figure 4: (a) Linear and, (b) semi-logarithmic plots of the dispersion relation for a parallel plate waveguide at optical frequencies. The dielectric constant $\varepsilon_m$ is that of a free-electron metal with a plasma frequency corresponding to $\hbar\omega_p=4$eV. The parameter d is the plate spacing.

The parallel plate inductance/unit length $L'_{cF}=\mu_o d/W$, and capacitance/unit length $C'_c=\varepsilon_o W/d$ are easy to derive, and well[12] documented. The cross plate capacitance $C'_c$ does not tell the whole story. We learned, when analyzing the single plate case that intra-plate capacitance, now labeled as $C'_i=2\varepsilon_o kW$ is also present. Likewise the intra-plate inductance $L'_{iF}=\mu_o/2kW$ must also be present.

Some corrections must now be introduced:

(*i*) Since the intra-plate inductance $L'_{iF}$ appears in series on both the upper plate and the lower plate, the correct value must be multiplied by 2×: $L'_{iF}=\mu_o/kW$. Like-wise the intra-plate capacitances on the upper and lower plates appear as reactive impedances in series. Thus the true intra-plate capacitance must be cut in half: $C'_i=\varepsilon_o kW$.

(*ii*) Further corrections are needed on the cross-plate inductance and capacitance/unit length. When the plate spacing is larger than the modal wavelength, only a fraction of the electric field lines reach from plate to plate, while the rest contribute to intra-plate capacitance. The presence of a spatially oscillating positive and negative charge on one plate implies that a distant plate sees net cancellation, a weak field that falls off exponentially as $\exp\{-kd\}$. Since the corresponding charge is smaller, for the same plate voltage, the cross-plate capacitance is smaller by the same factor, diminished to $C'_c=\varepsilon_o(W/d)\exp\{-kd\}$.

This exponentially decaying term is similar to the screening of electric field through a periodic perforated screen. The spatially oscillating surface charge provides the periodicity. Combining the intra-plate capacitance, $C'_i$, and the cross-plate capacitance, $C'_c$,

$$C' = C'_i + C'_c = \frac{\varepsilon_o W}{d}[kd + \exp\{-kd\}] \quad \text{............................(13)}$$

In the limit of small plate spacing $\exp\{-kd\} \to 1-kd$, and Eq'n. (13) reduces to that of a parallel plate capacitor, $C' \to \varepsilon_o(W/d)$. In the opposite limit of large plate spacing, $\exp\{-kd\} \to 0$, and $C' \to \varepsilon_o kW$, half the intra-plate capacitance, owing to the fact that the two widely separated plates are effectively in series.

Likewise the cross-plate inductance $L'_{cF}$ is increased when the plates are widely spaced, since little cross-plate displacement current flows in spite of a high voltage on the plate; $L'_{cF} = \mu_o(d/W)\exp\{kd\}$. This must be combined with the intra-plate inductance that we already know, $L'_{iF} = \mu_o/kW$. Since the current flowing in the metal plate flows either cross-plate or intra-plate the two inductance contributions, $L'_{cF}$ and $L'_{iF}$ must be in parallel:

$$L'_F = \left(\frac{1}{L'_{iF}} + \frac{1}{L'_{cF}}\right)^{-1} = \left(\frac{Wkd}{\mu_o d} + \frac{W}{\mu_o d \exp\{kd\}}\right)^{-1} = \frac{\mu_o d}{W[kd + \exp\{-kd\}]} \quad \text{............(14)}$$

In the limit of small plate spacing $\exp\{-kd\} \to 1-kd$, and Eq'n. (14) reduces to that of a parallel plate inductor, $L' \to \mu_o(d/W)$. In the opposite limit of large plate spacing, $\exp\{-kd\} \to 0$, and $L' \to \mu_o/kW$, twice the intra-plate inductance, owing to the fact that the two distant plates are in series.

The agreement between the exact model, Eq'n. (12), and the circuit model, insets of Fig. 4, is perfect in the limits of high and low wave-vector k, but there is some discrepancy at the knee of the dispersion. This may indicate of a more distributed interaction between the Faraday and kinetic inductance than was captured by our simple circuit model. As the plate spacing is increased, the fields at each metal plate gradually decouple from each other until the limit of a single surface wave guided by a single metal plate governed by Eqn. (6) and shown in Fig. (2).

The distributed incremental equivalent circuit for a small section of plasmonic parallel plate waveguide is illustrated in Fig. 5(a) and in the insets to Fig. (4), and includes both cross-plate and intra-plate circuit components. The single plate case is in Fig. 5(b) and the inset of Fig. (2), and includes only intra-plate circuit components. Figs. 5(a)&(b) also sketch the electric field lines and surface charges for a half-wavelength $\lambda_k/2$ segment of the line, at the wave-vector k. Fig. 5(c) operates at low enough frequency so that there are no plasmonic effects, i.e. negligible kinetic inductance, yet there remains an interesting competition between cross-plate and intra-plate fields. This case of intermediate wave-vector $(1/d)<k<(\omega_p/c)$ of Fig. 5(c) is non-plasmonic, yet it doesn't seem to appear in microwave text-books. For comparison the common case treated in textbooks is given in Fig. 5(d).

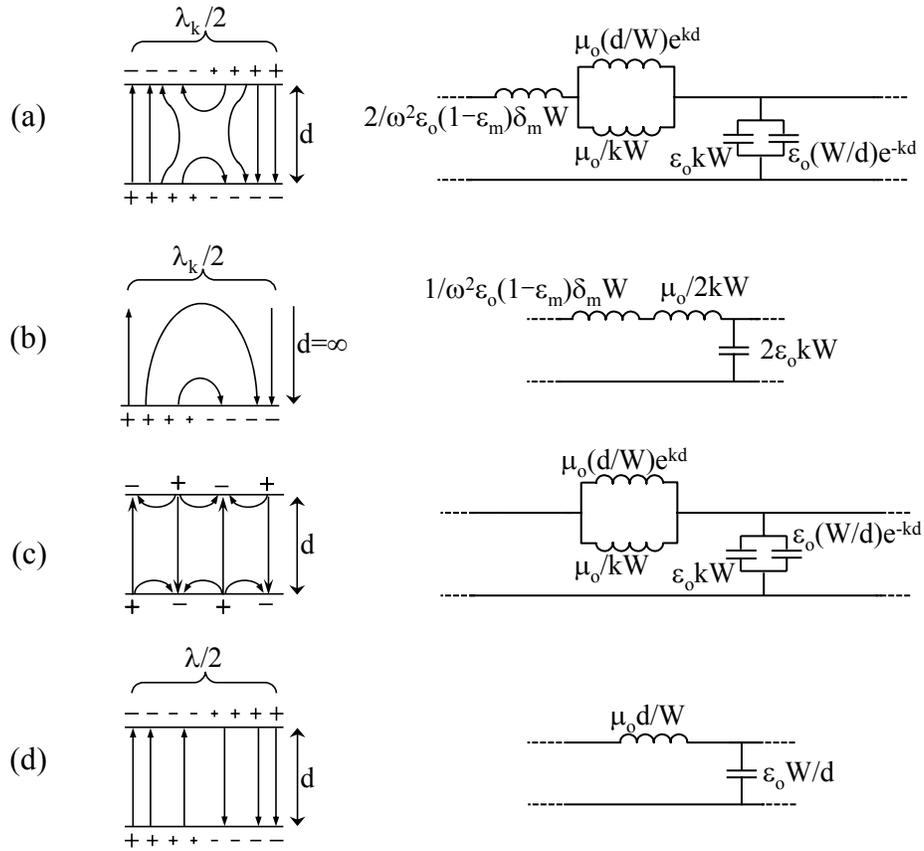

Figure 5: The electric field distribution, surface charges, and distributed equivalent circuits for (a) a half-wavelength $\lambda_k/2$ section of plasmonic parallel plate waveguide, of plate spacing d, and width W. (b) A half-wavelength $\lambda_k/2$ of a single-surface plasmonic plate of width W. (c) A two-wavelength section of parallel plate waveguide, at intermediate k along the light line, but having widely spaced plates: $(1/d)<k<(\omega_p/c)$. (d) A conventional RF parallel plate waveguide with $k<(\omega_p/c)$, and $kd<1$, as is usually covered in Electromagnetics texts[12]. Note that in all cases, when losses are present, the fields will additionally be bowed in the propagation direction[11].

As in any transmission line, the dispersion of Fig. 5(c) is given by $\omega^2=k^2/L'C'$. Substituting in the capacitance C' of Eq'n. (13), and the inductance L' of Eq'n. (14), all cross-plate and intra-plate terms cancel, leading to the simplest possible dispersion $\omega=kc$ along the light line. This dispersion in Fig. 5(c) is the same as an ordinary transmission line Fig. 5(d). The difference between Figs. 5(c)&(d) lies in the wave impedance $Z=\sqrt{(L'/C')}$. For the ordinary transmission line, Fig. 5(d), the wave impedance is controlled by the aspect ratio: $Z=(d/W)\sqrt{(\mu_o/\varepsilon_o)}$. For the widely spaced transmission plates of Fig. 5(c), the role of the spacing d is replaced by the reciprocal wave vector $Z=(1/kW)\sqrt{(\mu_o/\varepsilon_o)}$. Thus the widely spaced parallel plate waveguide has a wave-impedance even lower than a conventional waveguide, in the non-plasmonic regime, with kinetic inductance absent. In either Fig. 5(c) or (d), the wave impedance is always $<<377\Omega$, the impedance $\sqrt{(\mu_o/\varepsilon_o)}$ of free space.

*VI. Wave Impedance of the Plasmonic Parallel Plate Waveguide:*

In Equation (11) we have already given the surprising wave impedance of a single plasmonic plate, $Z=(\lambda_p/2\pi W)\sqrt{(\mu_o/2\varepsilon_o)}$, where $(\lambda_p/2\pi)\sim25nm$ is the collisionless skin depth. Uniquely, the wave impedance of a single narrow plate can become larger than $377\Omega$, the impedance of free space, owing to the additional contribution by kinetic inductance, $L_k$. Similar effects occur for the plasmonic parallel plate waveguide.

The plasmonic parallel plate wave impedance is $Z=\sqrt{[(L'_k+L'_F)/C']}$, where $L'_k = 2/\omega^2\varepsilon_o(1-\varepsilon_m)\delta_m W$, and C' & $L'_F$ are given by Eq'ns. (13)&(14) respectively. We are particularly interested in the regime where $L'_k$ makes a significant contribution. When $k>\omega_p/c$, and $k>1/d$, the intra-plate impedances dominate. The expression for Z simplifies to $(\lambda_p/2\pi W)\sqrt{(2\mu_o/\varepsilon_o)}$. This is twice the single plate impedance, as expected. This provides an opportunity for increasing the wave impedance above 377Ω.

When $k>\omega_p/c$, but k still less than 1/d, the following formula for wave impedance emerges:

$$Z = \sqrt{\frac{\mu_o}{\varepsilon_o}}\sqrt{\frac{d^2}{W^2}+\frac{2\lambda_p^2 kd}{(2\pi W)^2}}$$

which can be high, but not as high as $(\lambda_p/2\pi W)\sqrt{(2\mu_o/\varepsilon_o)}$. In either instance, kd>1 or kd<1, it is possible to achieve a wave impedance above 377Ω.

The ability to taper W to a narrower waveguide, provides in effect a transformer action at optical frequencies, for both single plate and parallel plate waveguides. Thus the optical ac voltage increases, and the optical ac current decreases, near the sharp tip. A decreased current, as a result of transformer action, is accompanied by diminished $I^2R$ resistive losses, which are the major problem in metal optics.

The availability of an optical voltage transformer suggests that efficient optical power delivery to the nanoscale is within[15] reach. The optical voltage transformer could also be used as a replacement for near-field scanning optical microscope (NSOM) probes[16] and as a heating element for heat assisted magnetic recording[17] (HAMR). When combined with optical antennas the optical voltage transformer could be used as an impedance matching tool to mediate molecules at high impedance, and antennas at a radiation resistance ~50Ω. This antenna matching could result in spontaneous hyper-emission[18], and contribute toward[19] surface enhanced Raman scattering (SERS).

*Appendix A: Derivation of capacitance per unit length for a wave on a flat metal plate:*

Consider a 2D charge distribution $\sigma_s$ in the plane, as shown in Fig. 3(a):

$$\sigma_s = \sigma_o \cos(kx), \quad \{ x \in (-\infty, \infty), \ y \in (-W/2, W/2) \}$$

The potential $V_o$ at the origin with respect to infinity is given by the usual integral over charge density:

$$V_o = \frac{1}{4\pi\varepsilon_o} \int_{-\infty}^{\infty} dx \int_{-W/2}^{W/2} dy \frac{\sigma_s}{\sqrt{x^2+y^2}} = \frac{1}{4\pi\varepsilon_o} \int_{-\infty}^{\infty} dx \int_{-W/2}^{W/2} dy \frac{\sigma_o \cos(kx)}{\sqrt{x^2+y^2}} \quad \ldots(A1)$$

The integration with respect to y may be evaluated using an indefinite integral from tables[20]:

$$\int_{-W/2}^{W/2} dy \frac{1}{\sqrt{x^2+y^2}} = 2 \int_0^{W/2} dy \frac{1}{\sqrt{x^2+y^2}} = 2\left[\ln\left|y+\sqrt{y^2+x^2}\right|\right]_{y=0}^{y=W/2} = 2\ln\left|\frac{W}{2}+\sqrt{\frac{W^2}{4}+x^2}\right| - 2\ln|x|$$

Inserting this integral into Eq'n. (A1) results in a strongly oscillating term, multiplied by a logarithmic function of x:

$$V_o = \frac{1}{4\pi\varepsilon_o} \int_{-\infty}^{\infty} dx \ \sigma_o \cos(kx) \cdot 2\left(\ln\left|\frac{W}{2}+\sqrt{\frac{W^2}{4}+x^2}\right| - \ln|x|\right) \quad \ldots(A2)$$

Changing variables to $kx \equiv \theta$, this becomes:

$$V_o = \frac{1}{4\pi\varepsilon_o k} \int_{-\infty}^{\infty} d\theta \ \sigma_o \cos\theta \cdot 2\left(\ln\left|\frac{kW}{2}+\sqrt{\frac{(kW)^2}{4}+\theta^2}\right| - \ln|\theta|\right) \quad \ldots(A3)$$

The condition for treating Fig. 3(a) as a 1-dimensional wave on a plane is that the wavelength should be much shorter than the width W of the plane. Then kW>>1, and the first logarithm in the integrand becomes simply ln|kW|, a constant number which multiplies the oscillating cosine, and averages to zero. . Then equation (A3) simplifies to:

$$V_o = \frac{-2\sigma_o}{4\pi\varepsilon_o k} \int_{-\infty}^{\infty} d\theta \cos\theta \cdot \ln|\theta| \quad\quad\quad\quad\quad\quad\quad\quad\quad\quad\quad\quad\quad (A4)$$

Eq'n. (A4) can be integrated by parts, with the integral converted to $-\int_{-\infty}^{\infty} (\sin\theta/\theta)\, d\theta$ which is equal[21] to $-\pi$. The peak voltage potential produced by all the surface charge is $V_o = \sigma_o/2\varepsilon_o k$. Since $V_o$ represents the peak value of a cosine, $V(x) = (\sigma_o/2\varepsilon_o k) \cos(kx)$. If we call C' the capacitance per unit length, then the surface charge per unit length becomes

$$C'V(x) = (C'\sigma_o/2\varepsilon_o k) \cos(kx) \quad\quad\quad\quad\quad\quad\quad\quad\quad\quad\quad\quad (A5)$$

We have an alternate expression for surface charge per unit length along the propagation direction, which is obtained by multiplying the surface charge density $\sigma_s$ times width W:

$$\sigma_s W = \sigma_o W \cos(kx) \quad\quad\quad\quad\quad\quad\quad\quad\quad\quad\quad\quad\quad (A6)$$

Requiring Eq'n. (A5) & (A6) to be equivalent, the capacitance per unit length along the propagation direction is: $C' = 2\varepsilon_o kW$.

*Appendix B: Derivation of inductance per unit length for a wave on a flat metal plate:*

As the surface charge shown in Fig. 3(a) oscillates in time, sinusoidal surface currents must flow in space. The surface currents produce a magnetic field B(x,z) above the surface, spatially sinusoidal in the x-direction. The effective Faraday inductance $L_F$ can be calculated[12] from $\int B\, dxdz = L_F I$, where I is the surface current over the full metal width W, and the magnetic flux is obtained by integrating $\int B dxdz$ above the metal surface in the +z-direction, and in the x-direction. Expressed as inductance/per unit length L' in the x-direction, the magnetic flux is

∫B dxdz=∫L'$_F$I dx.  Taking the integrands as equal simplifies this to L'$_F$=(∫B dz)/I.  We now show that L'$_F$=$\mu_o$/2kW.

In calculating B(x,y), it is helpful to use the vector potential A, just as it was helpful to calculate the scalar potential V in calculating capacitance.  Since all the currents flow in the x-direction, the only non-zero component is A$_x$:

$$A_x(z) = \int dx'dy \frac{\mu_o J_s}{r} = \frac{\mu_o}{4\pi} \int_{-\infty}^{\infty} dx' \int_{-W/2}^{W/2} dy \frac{J_s}{\sqrt{x'^2 + y^2 + z^2}} = \frac{\mu_o}{4\pi} \int_{-\infty}^{\infty} dx' \int_{-W/2}^{W/2} dy \frac{J_o \cos(kx')}{\sqrt{x'^2 + y^2 + z^2}} \ldots$$

..............(B1)

which falls off as 1/r from the source of current, just as scalar potential falls off as 1/r.  The distinction between x & x', is that x' is the variable of integration of the current density, and x is the variable of integration of magnetic flux.  The current density J$_s$=J$_o$cos(kx') is expressed per unit area, and integrated per unit area.

The magnetic field B can be derived from B=∇×A, but the only non-zero component is B$_y$=($\partial$A$_x$/$\partial$z).  Calculating the magnetic flux in the y-direction ∫($\partial$A$_x$/$\partial$z)dzdx and integrating only in z, the magnetic flux simplifies to $dx [A_x(z)]_0^\infty$.  An inspection of Eq'n. (B1) shows that A$_x$(z=∞)=0, but A$_x$(z=0) is finite.  This procedure is equivalent to using Stokes' Law for the magnetic flux:

$$\int B_y dxdz = \int (\nabla \times A_x) dxdz = \int A_x \cdot dl = A_x(z=0)\, dx$$

where the only part of the contour integral that is non-zero is along the incremental path dx, at the metal surface z=0.

Therefore the magnetic flux is A$_x$(z=0)dx, which by using Eq'n. (B1) can be written:

$$dx\, A_x(0) = dx \frac{\mu_o}{4\pi} \int_{-\infty}^{\infty} dx' \int_{-W/2}^{W/2} dy \frac{J_o \cos(kx')}{\sqrt{x'^2 + y^2}} \qquad \ldots\ldots\ldots\ldots\ldots\ldots\ldots\ldots\ldots\ldots(B2)$$

Eq'n. (B2) is identical in structure to Eq'n. (A1). The integration over y is an indefinite integral, and the integration over x' is a definite integral, as before. The integral reduces to:

$$dx\ A_x(0) = dx \frac{(-2)\mu_o J_o}{4\pi k} \int_{-\infty}^{\infty} d\theta\ \cos\theta \cdot \ln|\theta| = dx \frac{2\pi \mu_o J_o}{4\pi k} = dx \frac{\mu_o J_o}{2k} \quad \ldots\ldots\ldots\ldots(B3)$$

Since Eq'n. (B3) represents flux, then dx $A_x(0)$=LI=L'dx I, where I is the total current in the sheet $J_oW$. Then dx $(\mu_o J_o/2k)$=L'dx JoW. Cancelling equal terms on both sides of this equation, $L'_F = \mu_o/2kW$.

---

[1] N. Engheta, A. Salandrino, A. Alù, "Circuit elements at optical frequencies: nanoinductors, nanocapacitors, and nanoresistors", PRL 95, 095504 (2005)

[2] N. Engheta, "Circuits with light at nanoscales: Optical nanocircuits inspired by metamaterials", Science 317, 1698 (2007).

[3] A. Ishikawa, T. Tanaka, S. Kawata, "Negative magnetic permeability in the visible light region", PRL 95, 237401 (2005).

[4] J. Zhou, T. Koschny, M. Kafesaki, E.N. Economou, J.B. Pendry, C.M. Soukoulis, " Saturation of the Magnetic Response of Split-Ring Resonators at Optical Frequencies", PRL 95, 223902 (2005).

[5] N.A. Krall & A.W. Trivelpiece, *Principles of Plasma Physics*, McGraw-Hill, (1973). See page 123. The collisionless skin depth is defined as: $c/\omega\sqrt{(1-\varepsilon'_m)}$.

[6] S. Vedantam, M. Staffaroni, J. Conway & E. Yablonovitch, to be published.

[7] H. Raether, *Surface plasmons on smooth and rough surfaces and on gratings*, Springer Tracts in Modern Physics, Vol. 111 (1998)

[8] D.A. Cardwell, *Handbook of Super Conducting Materials*, CRC Press, (2003)

[9] E.N. Economou, "SurfacePlasmons in Thin Films", Physical Review Vol. 182, No. 2, 539-554 (1969)

[10] S.A. Maier, *Plasmonics: Fundamentals and Applications*, Springer, (2007)

[11] H.M. Barlow & A.L. Cullen, "Surface Waves", Proc. IEE 100, 399, (1953).

[12] F.T. Ulaby, *Fundamentals of Applied Electromagnetics*, Prentice Hall, (2007)

[13] J.D. Jackson, *Classical Electrodynamics*, Wiley, (1999)

[14] P. Ramo, J.R. Whinnery, & Van Duzer, *Fields and Waves in Communication Electronics, 3$^{rd}$ ed.,* Wiley (1994)

[15] M. Staffaroni et al, to be published.

[16] J. Tang et al, to be published.

[17] W.A. Challener & A.V. Itagi, *Near-Field Optics for Heat-Assisted Magnetic Recording (Experiment, Theory, and Modeling)*, Springer, (2009)

[18] E. Yablonovitch et al, to be published

[19] N. Kumar et al, to be published.

[20] H.B. Dwight, *Tables of Integrals and Other Mathematical Data, 4$^{th}$ Ed.*, MacMillan, Eq'n. No. 200.01 (1955).

[21] H.B. Dwight, *Tables of Integrals and Other Mathematical Data, 4$^{th}$ Ed.*, MacMillan, Eq'n. No. 858.601 (1955).